\documentclass[12pt,a4paper]{article}
\usepackage{graphicx}
\usepackage{amssymb}
\usepackage{amsmath}
\usepackage[british]{babel}

\newcommand{\AsAs}{A\&A}

\newcommand{\mnras}{MNRAS}

\newcommand{\nat}{Nature}
\newcommand{\apj}{ApJ}
\newcommand{\prd}{PRD}
\title{The CMB polarization: status and prospects}

\begin{document}

\author{Amedeo Balbi\footnote{amedeo.balbi@roma2.infn.it}, 
Paolo Natoli\footnote{paolo.natoli@roma2.infn.it}, 
and Nicola Vittorio\footnote{nicola.vittorio@roma2.infn.it}\\
{\footnotesize Dipartimento di Fisica, Universit\`a di Roma ``Tor
Vergata''}\\ {\footnotesize and INFN, Sezione di Roma ``Tor
Vergata''}\\ {\footnotesize Via della Ricerca Scientifica 1, I-00133
Roma, Italy}}

\date{}
\maketitle

\begin{abstract}
The study of cosmic microwave background (CMB) polarization is still
in a pioneering stage, but promises to bring a
huge advancement in cosmology in the near future, just as high-accuracy observations of the anisotropies in the total intensity of the CMB revolutionized our understanding of the universe in the past few years. In this
contribution, we outline the scientific case for observing CMB
polarization, and review the current observational status and future
experimental prospects in the field.
\end{abstract}

\section{Introduction}
Strong theoretical arguments suggest the presence of fluctuations in
the polarized component of the cosmic microwave background (CMB) at a
level of 5-10\% of the temperature anisotropy. A wealth of scientific
information is also expected to be encoded in this polarized signal.
However, while the existence of anisotropies in the temperature of the
Cosmic Microwave Background (CMB) has now been firmly established and
accurately measured by several experiments \cite{boom, max, dasi,
wmap}, bringing us high-precision constraints on the parameters of the
standard cosmological model, the investigation of the polarized
component of CMB anisotropy is still in its infancy.

It is then interesting to review, although briefly, the
 status of CMB polarization research, both theoretically and observationally. CMB polarization theory is well developed, while observations are complicated and hard to perform. In this contribution, we attempt to give a coverage of both issues: the basics of CMB polarization theory are outlined in Section~\ref{theory}, while the current and future observational situation is described in Section~\ref{observations}.

\section{Theory of CMB polarization}\label{theory}

In this section, we outline the basic theoretical framework of CMB
polarization, emphasizing the main aspects that make it an appealing
target for cosmological investigation. Excellent reviews on the
physics of the CMB polarization, exploring the subject with greater
detail, are \cite{Cabella & Kamionkowski, koso, primer}.

\subsection{Formalism}

The polarization properties of the CMB can be best understood using
the formalism of Stokes parameters \cite{chandra}. Consider a nearly
monochromatic electromagnetic wave propagating in the $z$ direction,
with $x$ and $y$ components of the electric field given by:
\begin{eqnarray}
  E_x&=&a_x(t)\cos\left[\omega_0 t - \theta_x(t)\right],\nonumber\\
  E_y&=&a_y(t)\cos\left[\omega_0 t - \theta_y(t)\right].
\end{eqnarray}
Then, the Stokes parameters are defined by:
\begin{eqnarray}
  I&\equiv& \langle a_x^2\rangle + \langle a_y^2\rangle,\nonumber\\
  Q&\equiv& \langle a_x^2\rangle - \langle a_y^2\rangle,\nonumber\\
  U&\equiv& \langle 2a_xa_y\cos(\theta_x -\theta_y)\rangle,\nonumber\\
  V&\equiv& \langle 2a_xa_y\sin(\theta_x -\theta_y)\rangle.
\end{eqnarray}
where the brackets $\langle\rangle$ represent the average over a time
much longer than the period of the wave.  The physical interpretation
of the Stokes parameters is straightforward. The parameter $I$ is
simply the average intensity of the radiation. The polarization
properties of the wave are described by the remaining parameters: $Q$
and $U$ describe linear polarization, while $V$ describes circular
polarization.  Unpolarized radiation (or natural light) is
characterized by having $Q=U=V=0$.  CMB polarization is produced
through Thomson scattering (see below) which, by symmetry, cannot
generate circular polarization. Then, $V=0$ always for CMB
polarization.  The amount of linear polarization along the $x$ or $y$
directions is measured by the Stokes parameter $Q$. When $U=0$, a
positive $Q$ describes a wave with polarization oriented along the $x$
axis, while a negative $Q$ describes a wave with polarization oriented
along the $y$ axis. Similarly, the Stokes parameter $U$ measures the
amount of linear polarization along the two directions forming an
angle of $45^\circ$ with the $x$ and $y$ axis. This implies that the
Stokes parameters $Q$ and $U$ are not scalar quantities. It is
straightforward to show that when the reference frame rotates of an
angle $\phi$ around the direction of observation, $Q$ and $U$
transform as:
\begin{equation}
\left(\begin{array}{c}
Q' \\ U'
\end{array}\right) = 
\left(\begin{array}{rr}
   \cos 2\phi & \sin 2\phi \\
  -\sin 2\phi & \cos 2\phi 
\end{array}
\right)
\left(\begin{array}{c}
Q \\ U
\end{array}
\right)
\end{equation}
This means that $Q$ and $U$ are not the components of a vector.
Mathematically, $Q$ and $U$ are the components of a second-rank
symmetric trace-free tensor:
\begin{equation}
  P_{ab}= \left( \begin{array}{rr}
   Q & U  \\
   U & -Q
   \end{array} \right),
\end{equation}
which represents a spin-2 field.  For visualization purposes only, one
can define a polarization ``vector'' with amplitude
$P=\left(Q^2+U^2\right)^{1/2}$ and orientation $\alpha=1/2
\tan^{-1}\left({U/Q}\right)$: this is not properly a vector, since it
remains identical after a rotation of $\pi$\/ around $z$, thus
defining an orientation but not a direction.

\subsection{Physical mechanism}

Having introduced the proper formalism, we can focus on the physical
mechanism that generates the polarized component of CMB radiation.
Before the formation of neutral hydrogen atoms in the primordial
universe (a process termed {\em recombination} which takes place a few
hundred thousand years after the big bang, at redshift $\approx
1100$), the CMB photons closely interact with the free electrons of
the primeval plasma through Thomson scattering. The angular dependence
of the scattering cross-section is given by:
\begin{equation}
{d\sigma\over d\Omega}={3\sigma_T\over 8\pi} \left\vert
   \hat\epsilon\cdot\hat\epsilon'\right\vert^2,
\end{equation}
where $\hat\epsilon$ and $\hat\epsilon'$ are the polarization
directions of incident and scattered waves.
After scattering, initially unpolarized light has: 
\begin{equation}
  I={3\sigma_T\over 16\pi}I'\left(1+\cos^2\theta\right), \ \ \ \ \ \
  Q={3\sigma_T\over 16\pi}I'\sin^2\theta, \ \ \ \ \ \
  U=0.
\end{equation}
Decomposing the incident radiation in spherical harmonics and
integrating over all incoming directions gives:
\begin{equation}
  I={3\sigma_T\over 16\pi}
  \left[{8\over 3}\sqrt{\pi}\, a_{00} + {4\over 3}\sqrt{\pi\over 5} a_{20}
  \right], \ \ \ \ \ \
  Q-iU={3\sigma_T\over 4\pi} \sqrt{2\pi\over 15} a_{22}.
\end{equation}
Then, polarization is only generated when a quadrupolar anisotropy in
the incident light at last scattering is present. This has two
important consequences. Because it is generated by a causal process,
CMB polarization peaks at scales smaller than the horizon at last
scattering. Moreover, the degree of polarization depends on the
thickness of last scattering surface. As a result, the polarized
signal for standard models at angular scales of tens of arcminutes is
about $10\%$ of the total intensity (even less at larger scales).
Typically, this means a polarized signal of a few $\mu$K.

\begin{figure}[!t]
\centering
  \includegraphics[width=\columnwidth]{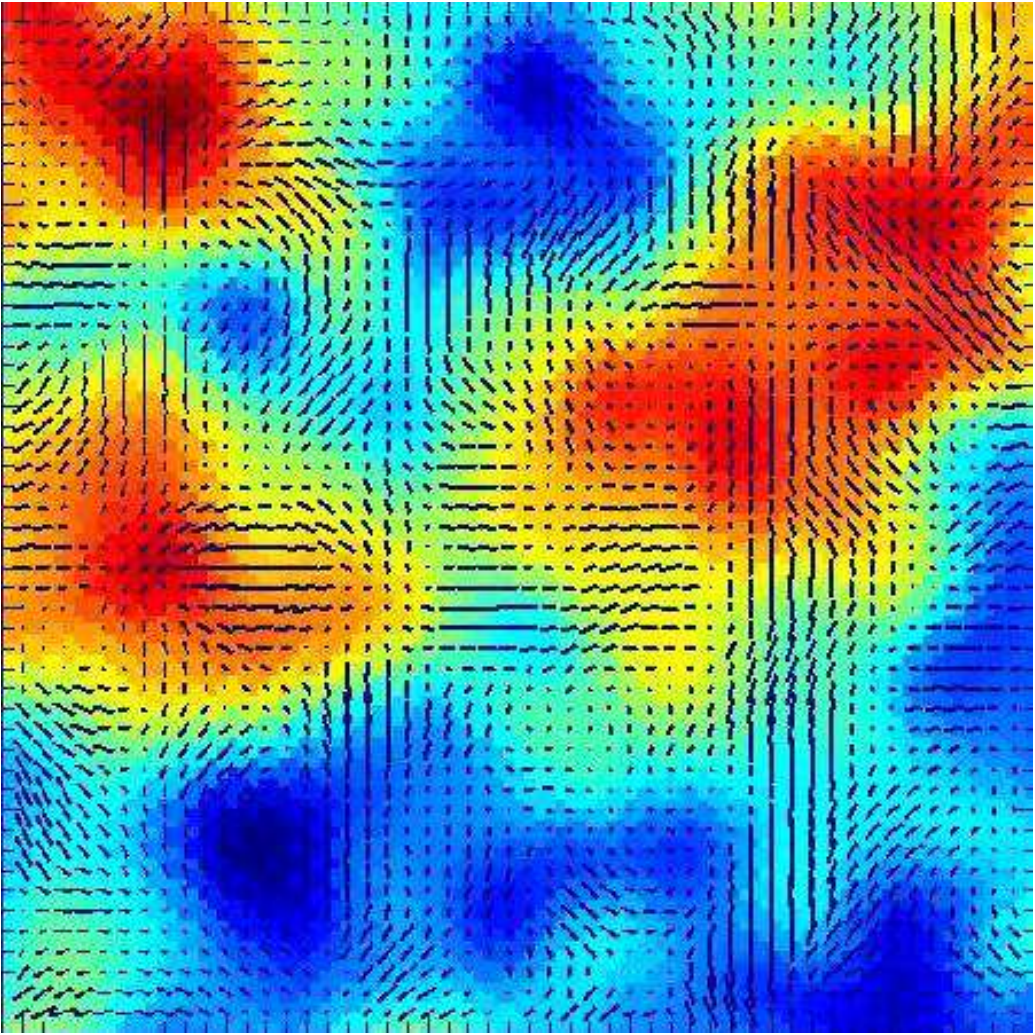} \caption{A
  simulated CMB temperature map and the corresponding polarized
  component represented by the polarization vector $P$, for a standard
  cosmological model where only scalar density perturbations are
  present.  The field is $6^\circ\times 6^\circ$, the resolution is
  $10^\prime$ FWHM.}\label{polamap}
\end{figure}

\subsection{Statistics}

\begin{figure}[!t]
  \includegraphics[width=\textwidth]{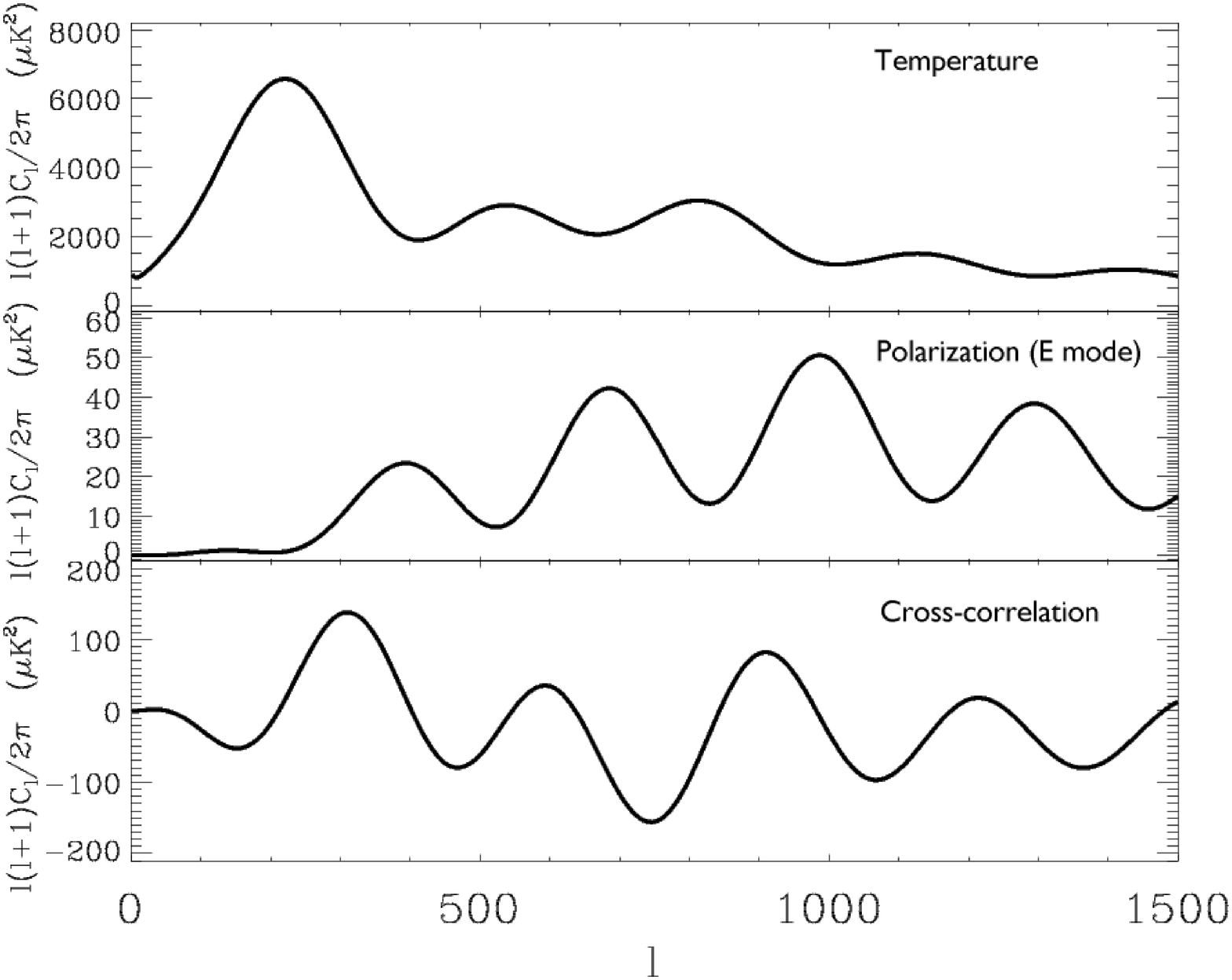} \caption{An example of
  the theoretical prediction for the power spectrum of CMB temperature
  fluctuations (upper panel), polarization E-mode (central panel), and
  cross-correlation between temperature and E-modes (bottom panel). It
  is quite noticeable the fact (described in the text) that peaks in E
  spectrum are out of phase with those in the T spectrum. Polarization
  also peaks on smaller angular scales and has a much lower intensity
  than temperature.}\label{spectra}
\end{figure}

In order to connect the observed polarized signal to theoretical
prediction, it is convenient to expand the polarization tensor in
terms of two scalar fields: a so-called electric (or gradient) field,
$E$, and a magnetic (or curl) field, $B$. This is analogous to the
decomposition of a vector into a gradient and divergence-free
vector. These fields can then be expanded into spherical harmonics
just as it is done with the temperature field, as:
\begin{equation}
  {\bf P}_{ab} =  T_0\sum_{l=2}^\infty\sum_{m=-l}^l \left[
    a_{(lm)}^{\rm E} Y_{(lm)ab}^{\rm E} + a_{(lm)}^{\rm B}
    Y_{(lm)ab}^{\rm B}  \right]
\end{equation}
where the generalization of spherical harmonics for arbitrary spin has
been used \cite{varsha}.  The statistical properties of the CMB
anisotropies polarization are then characterized by six power spectra:
$C_l^T$ for the temperature, $C_l^E$ for the E-type polarization,
$C_l^B$ for the B-type polarization, $C_l^{TE}$, $C_l^{TB}$,
$C_l^{EB}$ for the cross correlations. Each of the power spectra is
computed in the usual way from the spherical harmonic coefficients,
as:
\begin{equation}
\langle a^{X\star}_{lm}a^{X'}_{l'm'}\rangle=C^{XX'}_l\delta_{ll'}\delta_{mm'}
\end{equation}
For the CMB, $C_l^{TB}=C_l^{EB}=0$. Furthermore, since $B$ relates to
the component of the polarization field which possesses a handedness,
one has $C_l^{B}=0$ for scalar density perturbations. The detection of
a non zero $B$ component would then point to the existence of a tensor
contribution to density perturbations. 
An example of theoretical power spectra is shown in Figure~\ref{spectra}

The relation relating ($E$,$B$) to ($Q$,$U$) has a non-local nature.
This can be easily seen in the limit of small angles, where it can be written as:
\begin{eqnarray}
  E(\bf{\theta})=-\int d^2{\bf \theta}^{\prime}\
  \omega(\tilde \theta)\ Q_r({\bf \theta}^{\prime}), \nonumber\\  
  B(\bf{\theta})=-\int d^2{\bf \theta}^{\prime}\
  \omega(\tilde \theta)\ U_r({\bf \theta}^{\prime}),
\end{eqnarray}
where the 2D angle $\bf{\theta}$ defines a direction of observation in
the coordinate system perpendicular to $z$,
\begin{eqnarray}
  Q_r({\bf \theta})=Q({\bf \theta}^{\prime})
  \cos(2\tilde\phi) - U({\bf \theta}^{\prime})
  \sin(2\tilde\phi), \nonumber \\ 
  U_r({\bf \theta})=U({\bf \theta}^{\prime})
  \cos(2\tilde\phi) + Q({\bf \theta}^{\prime})
  \sin(2\tilde\phi).
\end{eqnarray}
and $\omega(\tilde \theta)$ is a generic window function.

\subsection{CMB polarization as a cosmological tool}

\begin{figure}[!t]
\centering
  \includegraphics[width=\textwidth]{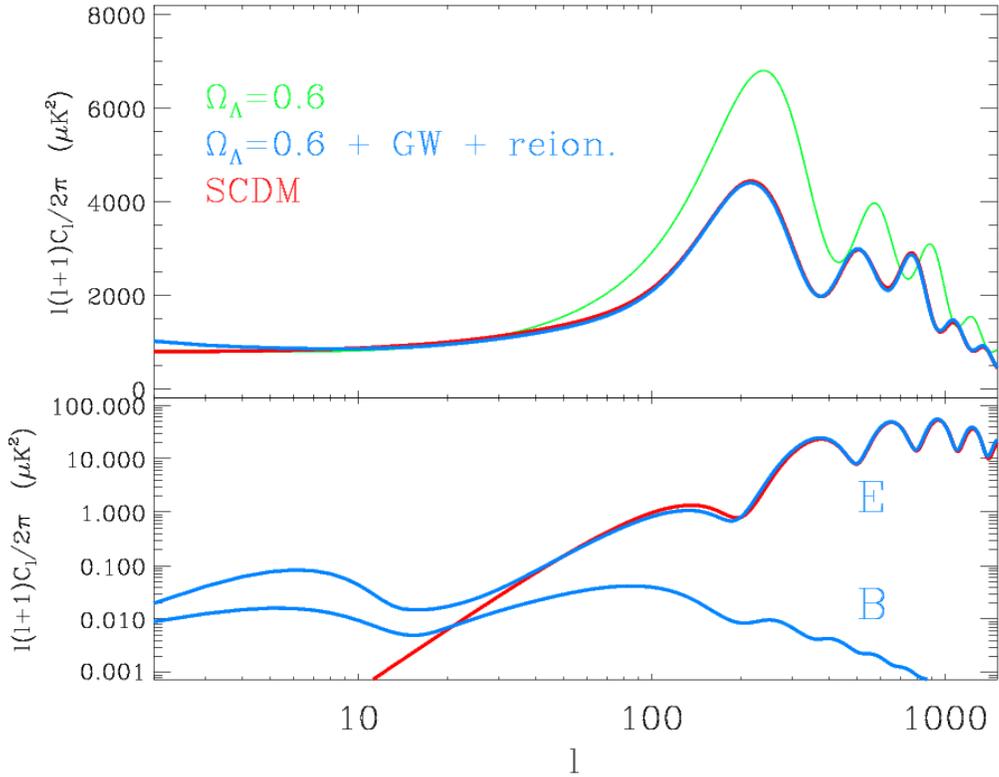}
  \caption{{\em Top panel\/} -- Temperature power spectra for a standard 
    CDM model (solid line), and the same CDM model with a fraction of
    the critical density coming from a cosmological constant (dotted
    line). If we add to the latter a contribution from tensor
    perturbations (gravitational waves background) and reionization
    (dashed line) we can make it indistinguishable from the standard
    CDM model. {\em Bottom panel\/} -- The same models and their
    polarization power spectra. The CDM model can be identified by its
    polarized signal, because it does not generate a B-type
    component}\label{polaps}
\end{figure}

The polarization power spectrum for the $E$ mode shows the same kind
of acoustic features that are now well known to exist in the
temperature anisotropy spectrum. The peaks in polarization, however,
are out of phase with those in the temperature: $E$ mode polarization
has maximum intensity were the temperature is at a minimum, and
viceversa. This is due to the fact that polarization is generated by
quadrupolar anisotropy at last scattering, and this is closely related
to the velocity of the coupled photon-barion fluid. The maximum
compression of rarefaction (and minimum velocity) of the fluid
corresponds to peak in the temperature anisotropy (and troughs in the
polarization). For the same reason, the cross-correlation power
spectrum between temperature and $E$ polarization shows pronounced
peaks corresponding to the interleaved sets of maxima and minima in
the two separate components. A detection of such features in the
polarization and cross-polarization power spectra is then be
crucial for two reasons: first, to give an independent
confirmation of the fact that the temperature anisotropy peaks are a
sign of genuine acoustic oscillation in the primeval plasma; second,
to confirm the adiabatic nature of primordial perturbations
(the only known way to produce acoustic oscillations) thus providing
strong support to cosmic inflation.

A large amount of valuable cosmological information can be extracted
from the observation of CMB polarization. Since it probes the epoch of
decoupling, polarization allows one to perform detailed tests of the
recombination physics. In particular, it is a well known fact that the
universe underwent a phase of reionization at redshifts of at least
$z\sim 5$, during the formation of early cosmic structure. The
investigation of this so-called dark ages is an active subject of
investigation \cite{Choudhury & Ferrara 2006}. The CMB temperature
anisotropy signal gets damped when the photons are diffused by free
electrons along the line of sight. The amount of damping would be a
powerful probe of the optical depth to reionization. However this
effect is masked by other physical mechanisms. For example, there is a
strong degeneration with the spectral index of primordial
perturbations. CMB polarization would prove very powerful in removing
these degeneracies: if the optical depth is non zero, a recognizable
polarization signature gets generated at large angular scales,
allowing to investigate the detailed reionization history,
discriminating models that have the same optical depth but a different
evolution of the ionization fraction with redshift \cite{Hu & Holder
2003}. Not only the characterization of the detailed ionization
history of the universe would have a strong scientific impact, but it
would also greatly increase the accuracy of the determination of other
cosmological parameters, such as the above mentioned spectral index of
scalar primordial perturbations. This, in turn, would be extremely
important when constraining models of inflation.

One crucial aspect of CMB polarization has to do with the properties
of its $B$ component. The consequences of detecting $B$ polarization
for theoretical models would be enormous. $B$ modes, as mentioned
above, can only be generated when a tensor component of primordial
perturbations is present (namely, a background of primordial
gravitational waves). This is precisely one of the prediction of
inflationary scenarios. The ratio of scalar to tensor fluctuations
$r$, related to the amplitude of $B$ modes, is a direct signature of
the inflation energy scale. On the other hand, observing the $B$
component is the hardest challenge faced by experimentalists, since
the signal is expected to be extremely faint, and possibly
contaminated by diffuse galactic emission. Furthermore, the $B$
spectrum should peak at large angular scales (the most affected by
cosmic variance) and can be contaminated by spurious leakage from $E$
modes when the observed area does not cover the entire sky. The level
of $B$ signal also depends very strongly on the epoch and amount of
reionization, so that the predictions of detectabilities are affected
by the assumptions made on the ionization history of the
universe. Finally, weak lensing from large scale structure in the
local universe affects the distribution of CMB photons (resulting in a
so-called cosmic shear). This can induce a curl component in the
polarization signature, that can be wrongly interpreted as evidence
for non zero $B$ modes, or act as a background that would make the
tensor primordial contribution impossible to detect \cite{Knox & Song
2002}. Techniques to detect and remove the shear contribution have
been developed and should prove effective when dealing with future
data \cite{Seljak & Zaldarriaga 1999}. The shear contribution has also
a scientific interest in itself, since it has been shown \cite{shear}
that it can lead to a determination of the neutrino mass from
measurements of CMB temperature and polarization alone.

\section{Detecting CMB polarization}\label{observations}

\subsection{Past and present}

\begin{figure}[!t]
  \resizebox{\hsize}{!}{\includegraphics[angle=0]{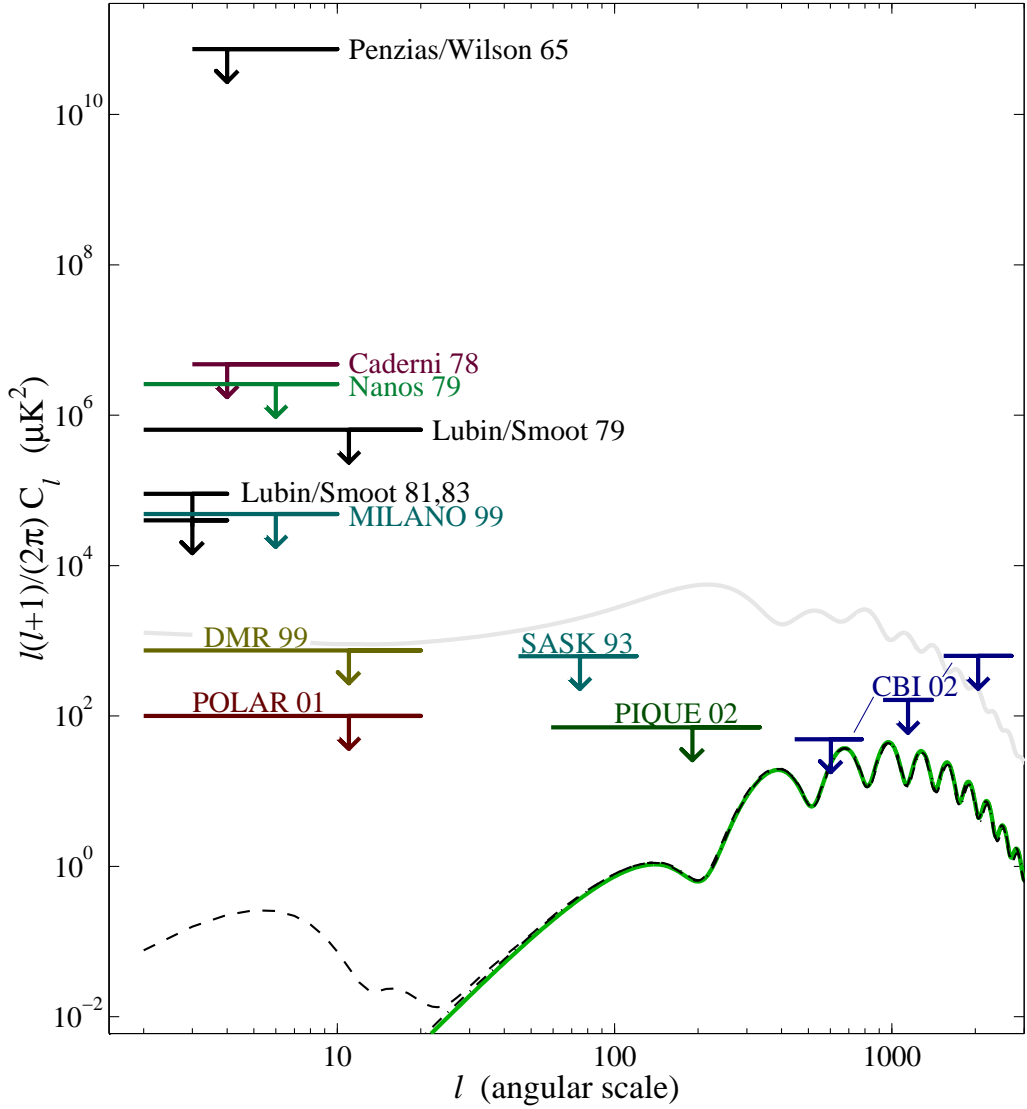}}
\caption{The situation in polarization
 experimental limits up to 2002. See text for references. Reprinted
 from \cite{2003NewAR..47..953C}.}
\label{experimental_limits}
\end{figure}

It has been realized as early as 1968~\cite{Rees} that, close
to decoupling, Thomson scattering would induce a small degree of
linear polarization on unpolarized but anisotropic CMB photons. However, the
weakness of this contribution has prevented its detection until very recent
times. Over three decades of experimental efforts have resulted in a
quite lengthy series of upper limits
(e.g.~\cite{1965ApJ...142..419P,1979ApJ...232..341N,Caderni et al,Lubin & Smoot,1981ApJ...245....1L,1983ApJ...273L..51L,1998NewA....3....1S,Partridge et al,1993ApJ...419L..49W,Netterfield et al,2003PhRvD..68h3003D})
some of which are displayed in Fig.~\ref{experimental_limits}.
These limits speak of a steady increase towards the level of the
polarization anisotropy, a quest well motivated by the understanding
--- that had become clear along road --- of the wealth of
cosmological information encoded in the polarization power spectra.

The long pioneering phase was ended by the
DASI~\cite{2002Natur.420..763L} detection of a TE and EE signal in
2002. DASI is a ground based interferometer located at the South Pole
Amundsen-Scott research station.  When configured as a polarimeter, it
has sensitivity to all four Stokes parameter, and has been optimized
to study CMB anisotropy in the range $140\lesssim\ell\lesssim 900$
\cite{2002Natur.420..772K}.  The first release has reported a
detection of EE mode polarization with an rms amplitude of 0.8
$\mu\mathrm{K}$ at 4.9 $\sigma$ and a $\sim 2 \sigma$ detection of TE,
values strengthened to 6.3 $\sigma$ (2.9 $\sigma$) for EE (TE) in the
more recent three years release~\cite{2005ApJ...624...10L}.

Following DASI, other experiments have claimed EE detection; CAPMAP,
an evolution of the PIQUE system~\cite{1997ApJ...476..440W}, using
coherent HEMT receivers at 100 GHz coupled to a large dish to achieve
high (3.6') angular resolution, has reported marginal ($2.3~\sigma$)
detection of EE~\cite{2005ApJ...619L.127B}. CBI (Cosmic Background
Imager), a radio interferometric receiver working in the band 26-36
GHz and covering a wide range of multipoles ($300\lesssim\ell\lesssim
3500$) has reported the most robust EE detection to date on small
angular scales, at 8.9~$\sigma$
\cite{2004Sci...306..836R}.
\begin{figure}[!t]
  \resizebox{\hsize}{!}{\includegraphics[angle=0]{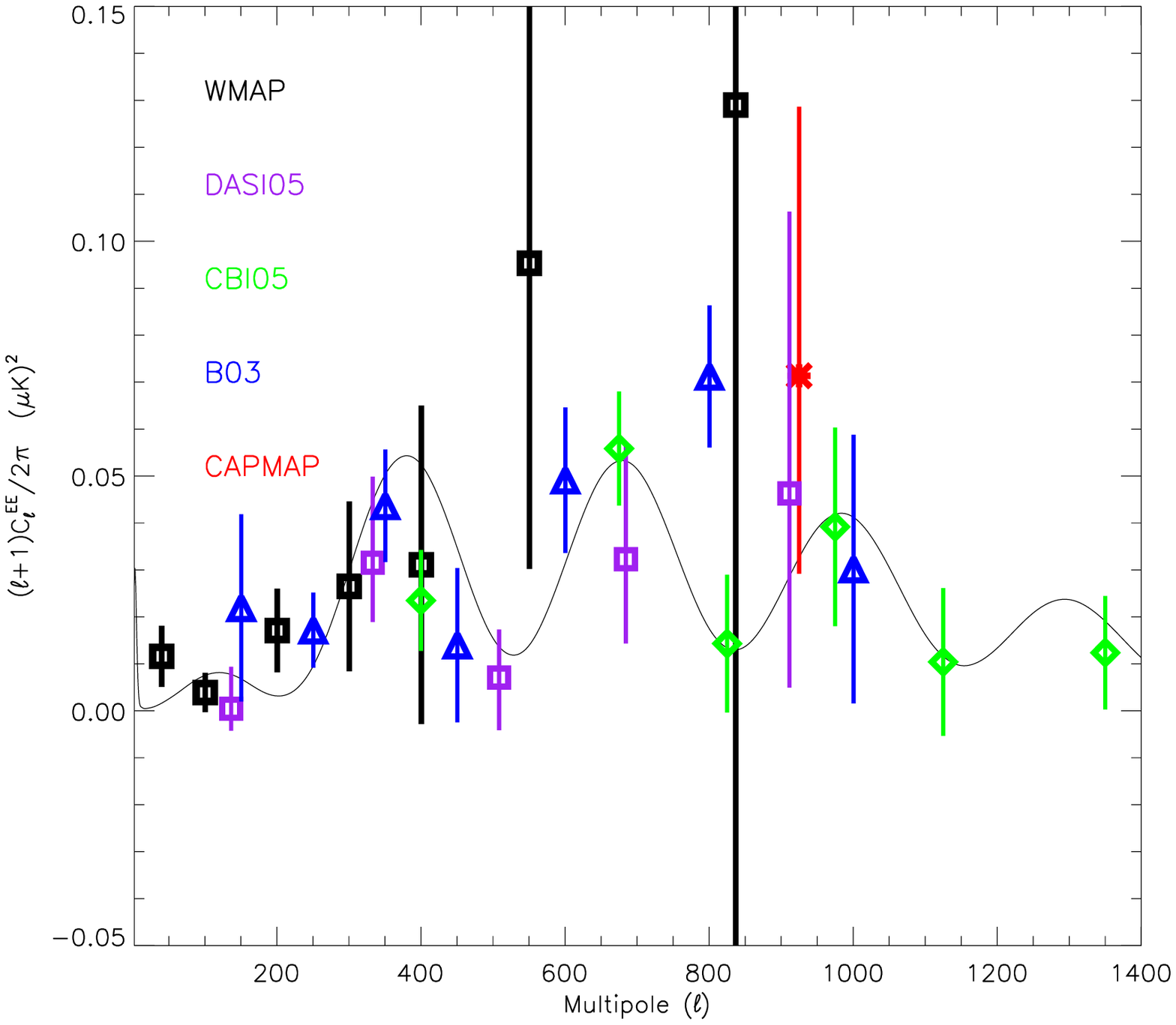}}
\caption{Current measurements of the EE mode
 power spectrum. Note that the band values are multiplied by $\ell+1$,
 not the more usual factor $\ell(\ell+1)$ . Reprinted
 from~\cite{wmap_yr3_pol}.}
\label{wmap_ee_comp}
\end{figure}
The Wilkinson Microwave Anisotropy Probe (WMAP) satellite has measured
TE correlations with high $\ell$ resolution at many $\sigma$ as of the
first data release of 2003~\cite{2003ApJS..148..161K}. As it is well
known, WMAP is a satellite flown by NASA featuring HEMT radiometers
arranged in differential assemblies; it observes the sky in four
microwave bands, K$_a$, Q, V and W. The three year WMAP
release~\cite{wmap_yr3_pol} has improved significantly its TE measure,
particularly at the lowest multipole probed, sensitive to the CMB
photon's optical depth $\tau$. Furthermore, it has provided an EE
detection, again especially effective at the (hitherto untested)
lowest multipoles, given the combination of the WMAP large sky
coverage and modest sensitivity (as far as CMB polarization is
concerned).

The above experiments have detected EE modes (and, in some cases, TE
correlations) by using coherent detectors. Measures of CMB
polarization using uncoherent detectors (namely, bolometers) have
been delivered by BOOMERanG 03, a balloon borne experiment featuring a
new generation of bolometers dubbed PSB (Polarization Sensitive
Bolometers,\cite{2003SPIE.4855..227J}), flown in 2003 from the
McMurdo Antarctic base.  Analysis of the flight data
\cite{masi_b2k,piacentini_b2k,montroy_b2k} have provided
high quality measurements of TE and EE CMB spectra. The current
situation of EE measurements is displayed in Fig.~\ref{wmap_ee_comp}.

\subsection{Current implications for cosmology}

The present status of CMB polarization observations does not allow a
determination of cosmological parameters at the same level of accuracy
obtainable using temperature anisotropies. The most useful
cosmological result established using CMB polarization data is,
currently, one of consistency: the standard cosmological model (a
universe with flat geometry, dominated by dark matter and dark energy
of unknown nature, and consistent with the basic predictions of the
inflationary scenario) predicts a level of CMB polarization that is
not ruled out by observations, and actually seems to be of the right
intensity to match the present data. Intriguingly, the peak and
through positions in the E polarization spectrum seems to fall in the
right places to match the predictions of the adiabatic primordial
perturbation scenario (see, e.g.\ \cite{MacTavish et al. 2005}. The
level of B polarization is well within the upper limits derived from
the observations. These results lend further support to the basic
cosmological scenario --- although they are undoubtedly in need of
stronger confirmations.

Perhaps, the strongest cosmological constraints from CMB polarization
are currently indirect ones. For example, when the WMAP EE data are
taken into account, they help to exclude a significant set of values
for the optical depth to reionization.  Claims of a very early
reionization epoch from the first year release of WMAP data
\cite{Spergel et al. 2003} were essentially driven by a large TE
correlation detected at large angular scales: this has now
dramatically changed, once better data from three years of
observations and new constraints on E modes were taken into account
\cite{Spergel et al. 2006}. Better constraints on $\tau$ have, in
turn, had a consequence for the determination of the primordial
spectral index, since the degeneracy between $\tau$ and $n_s$ got
significantly reduced.

To summarize, the CMB polarization looks currently very promising as a
tool to constrain cosmological models, but observations are not at the
level of accuracy needed to push the envelope of parameter
determination. Clearly, much better data are needed, particularly in
the area of B modes, still almost unexplored.


\subsection{Experimental concerns}

Before CMB polarization measurements can be turned into an 
high precision cosmological tool, thorough understanding of  
many experiment related issues is required. In comparison
with intensity anisotropy measurements, significant
complications arise. 

One obvious issue is detector sensitivity: detecting the CMB E mode
polarization requires an instrumental sensitivity a couple of orders
of magnitude higher than intensity, and even in the most favorable
models the B mode signal is a further order of magnitude below
E. However, very little room is left for improvement in single detector
technology: bolometers used in modern CMB missions (such as BOOMERanG
and \textsc{Planck} HFI) are already photon noise
limited~\cite{2003SPIE.4855..227J}, while coherent detectors (e.g.,
HEMT based radiometers) are not expected to improve significantly in
the near future\cite{2004AIPC..703..385B}. The only possible way to
increase sensitivity is thus by statistical redundancy, increasing the
total integration time and/or the number of detectors that populate the
focal plane. The former strategy clearly speaks in favor of ground
based and space borne missions, although Ultra Long Duration
Ballooning (ULDB) CMB aimed flights could be achieved in the near
future. In any case, long term stability of highly sensitive
instruments is a matter of concern (see below). On the other hand,
integration of many detectors on a single focal plane seems to be an
unavoidable complication. Concerns here mainly relate to manufacturing
costs, since it is unsuitable to mass produce highly sensitive CMB
detectors, and, in the case of space borne experiments, power budget
availability, since active cooling is likely to be required. In
addition, it is unclear how high density focal plane will behave in
terms of cross talk between detectors. The current trend is towards
photolithographic arrays of bolometers, although techniques are
being devised that could make radiometer arrays perfectly feasible
and competitive.

Detector sensitivity is not the full story.  Not surprisingly,
polarization measurements inherit and magnify all the contaminations
by systematic effects that have plagued high precision temperature
observation. In addition, being polarization a tensor quantity, its
measurement strategies are usually more complicated than those adopted
for intensity. Hence, a number of polarization specific systematic
effects have been predicted --- and encountered --- that are likely to
complicate or even hamper detection, regardless of statistical noise
control. The long list includes optics induced polarization (most
telescopes used for CMB are off-axis system due to the necessity of
controlling sidelobes), band and gain mismatch between detectors,
cross polarization (i.e.\ far from perfect isolation effects) and beam
asymmetry, that induces leakage of I into Q and U. Furthermore, due
to limiting scanning strategy constraints, polarization is usually
estimated by joint analysis of several detectors, both in order to
gain sensitivity and account for poor redundancy in detector
orientation.  This fact further complicates the analysis, as optical
beam mismatches and detector cross talks become sources of
concern. For instance, BOOMERanG 03 had to correct for noise
correlated among different detectors~\cite{masi_b2k}, while WMAP has
encountered problems related to gain
variations~\cite{2006astro.ph..3452J}. The above complications also
reflect onto data analysis techniques, which often must be redesigned
to tackle polarization (e.g., \cite{2005A&A...436.1159D}).
 
Our best window for observing the temperature anisotropies lies close
to 70 GHz, where foreground emission seems to display a minimum. This
has permitted accurate mapping of the CMB intensity pattern, without
sacrificing much sky to foregrounds. However, when precision
measurements are at the stake, cleaning or component separation
techniques are needed(e.g., \cite{2003PhRvD..68l3523T,2003NewAR..47.1127B}. \textit{A fortiori},
the same is deemed true for polarization measurements, although not
much is known about polarized foreground emission. Archeops has
measured polarized emission from dust at 353 GHz both in the Galactic
plane~\cite{2004A&A...424..571B} and at higher Galactic
latitudes~\cite{2005A&A...444..327P}. In the latter case,
extrapolation of their findings suggest that polarized dust emission
will be a major source of concern for measurements above 100
GHz~\cite{2005A&A...444..327P}. WMAP found significant contamination
from polarized foreground extended to large portions of the sky; the
WMAP team analysis shows that a foreground model must be subtracted
from the data to obtain meaningful detections of CMB polarization at
low ($\ell < 10$) multipoles and that the cleanest window for
polarization measurements is located at about 60
GHz~\cite{wmap_yr3_pol}.  Under the hypothesis that all other
statistical and systematic error sources will eventually be tamed,
foreground contaminations are likely to dominate the uncertainties in
future measurements of the CMB anisotropy pattern, just like what is
happening today with intensity. It is also likely that component
separation methods will play an important role in the analysis (e.g.,
\cite{2004astro.ph..3175D}).

\subsection{The near future}

Several experimental efforts to measure the polarization of the CMB
are en route. A few are currently in operation or in the (often
lengthy) process of data analysis, while the majority will are
expected to deliver data within the next few years. Hereafter, we list
a few experiments, loosely divided by type, without sake of
completeness. See~\cite{2004astro.ph..3175D} for a review.

\textit{Ground based experiments}--- Several E modes oriented efforts are 
still active, including a few mentioned above; however, the tendency
is towards the detection of B modes. AMiBA~\cite{2003AstHe..96..374U},
is a dual channel (85 -- 105 GHz) interferometric array of (up to) 19
elements, currently being installed at the Mauna Kea (Hawaii)
site. Its predicted angular resolution is about 2 arcminutes, and has
full polarization capabilities: contrary to most CMB oriented
projects, it can also measure the circular polarization Stokes
parameter ($V$). BICEP (Robinson Gravitational Wave Background
Telescope)
\cite{2006astro.ph..6278Y} is an array featuring 49 pairs
of polarization sensitive bolometers already in operation at South
Pole.  It has an instantaneous field of view of $\sim 17^\circ$, and
angular resolutions of $55'$ ($37'$) at 100 (150) GHz. BICEP has the
sensitivity to detect in three years of observations the peak induced
by primordial gravitational waves in the BB spectrum, if $r \gtrsim
0.1$.  BRAIN and ClOVER are two separate but interlinked (in the sense
that they rely on similar hardware) experiments to be deployed at the
Dome C observing site, within the French-Italian Concordia Antarctic
facility~\cite{2004sf2a.conf..707P}. Dome C is probably the best observing site
on Earth for high precision microwave measurements.
BRAIN~\cite{2005EAS....14...87M} is a new concept bolometric
interferometer using six pairs of coupled feed horns operating at 150
GHz and spaced by a few wavelength; BRAIN's pathfinder experiment is
under construction on site. ClOVER~\cite{2005EAS....14..251M} is a
multiple bolometric array: four arrays each composed of 64 bolometers
are at the focal planes of optical assemblies, evenly arranged around
the cryostat. There are a total of three telescopes in the range 90 --
220 GHz, its angular resolution for CMB oriented channels being about
$15'$. ClOVER is scheduled to achieve first light in 2008. QUaD (QUEST
at DASI)~\cite{2003NewAR..47.1083C} is another new generation
instrument created by mounting the QUEST bolometer array on the DASI
telescope. QUEST~\cite{2002AIPC..609..159P} is an array of 31
bolometers operating between 100 and 150 GHz. Observing up to $\ell
\sim 2500$, it can measure the contribution to B modes arising
from lensed E modes, i.e.\ the so called cosmic shear. The above
instruments are all based on bolometric technology. An example of an
experiment that employs an array of coherent detectors (``radiometers on a
chip'') is QUIET~\cite{2005AAS...20717015N}.

\textit{Balloon borne experiments}--- The data analysis from BOOMERanG 
is still ongoing and the balloon could in principle be flown
again. MAXIPOL~\cite{2003NewAR..47.1067J} is another balloon aimed at
E modes, featuring a rotating half wave plate in front of bolometric
detectors. It has been flown, and a data release is expected soon.
Bar-SPORT~\cite{2003SPIE.4843..324Z} is an experiment based on
coherent detectors to measure the E mode pattern, devised as a test
bed for SPORT (see below). Originally planned as a balloon borne
telescope, it now seems to have been reconfigured as a ground based
effort at Dome C~\cite{2005EAS....14..257Z}.  Among the experiments
also aimed at the B modes, EBEX~\cite{2004SPIE.5543..320O} is an
effort based on bolometric transition edge detectors; polarization
capabilities are achieved through a magnetically levitated halve
wave plate. The focal plane is shared between four arrays observing at
150, 250, 350 and 450 GHz, for a total of up to 1320 detector
elements, with a maximum resolution of $8'$. Another B modes oriented
balloon, funded by NASA, is PAPPA~\cite{erpappone}

\textit{Orbital efforts}--- WMAP is obviously still taking data, slowly
increasing its sensitivity with time. SPORT~\cite{2002ASPC..257..243C}
is an experiment selected by ESA to be flown on the International
Space Station. It is based on coherent detector and directly measures
the Q and U Stokes parameters (on a $7^\circ$ angular scale) by cross
correlating two circular components with opposite helicity. Originally
scheduled for launch in 2005, it has been delayed (currently,
\textit{sine die}) due to the well known NASA Space Shuttle program
difficulties.
\textsc{Planck}~\cite{planck}, by many considered the ultimate CMB 
temperature anisotropy mission, has polarization capabilities in all
its coherent (LFI) detectors and in many bolometric (HFI) detectors.
The launch is currently scheduled in early 2008.
The uniqueness of \textsc{Planck} is in that it will produce Q and U maps
over the full sky over a wide range of frequencies (30 -- 350 GHz), with the
combination of high sensitivity and tight control of systematics that
is only achievable from space. However, \textsc{Planck} has not been
designed with B modes in mind, and a mission with at least a sensitivity
10 times better is needed to detect $r \sim 0.01$.

\subsection{Exploiting the B modes: an experimental challenge}
 
It is quite evident that the current goal in CMB science is to measure
the B modes and distillate cosmological information out of them. The
two hot fronts of this research are the ``low'' $\ell$ $(\sim 100)$
signal from primordial gravitational waves and the high $\ell$
contribution from lensed E modes. It is likely that, if inflation is a
reliable theory and if there are no unforeseen systematic complication,
one or more sub orbital efforts among the ones listed above will
actually detect these modes. Another question, however, is whether
such a detection will prove robust and accurate enough to derive firm
consequences for Cosmology. It is a widely accepted argument that
only a satellite mission can achieve the correct combination of
full sky sampling, sensitivity, thermal stability and spectral
coverage that will turn the detection into solid science. In this
respect, the many suborbital efforts under development will act
as pathfinders, returning invaluable testing of frontier technological
devices.

Several post \textsc{Planck} orbital efforts are under study. NASA has
solicited proposals for its ``Beyond Einstein'' program in structure
and evolution of the universe; in particular, the request is for a
medium size mission code named Inflation Probe, to search for the
gravitational wave signature in the CMB. The current Inflation Probe
candidate is called CMBPOL~\cite{2006astro.ph..4101B}. If selected,
this mission could be flown as early as 2018. In Europe as well
concept studies are being submitted to national agencies, as is the
case of the French SAMPAM project~\cite{2005sf2a.conf..675B}. Within a
few months, ESA will probably issue a call for proposal for its Cosmic
Vision program. It is almost certain that a specific proposal will be
submitted by the CMB community for a B modes oriented satellite.
Whether this will be for a low $\ell$ or an high $\ell$ mission
(capable, e.g.\, to measure the mass of the neutrino) is still
unclear.

\section{Conclusions}

We have presented a brief overview of the status and prospects of CMB
polarization research. The scientific potential of investigating
polarization cannot be underestimated: we can obtain crucial
information on such topics as the physics of reionization, the energy
scale of inflation, we can get much better constraints on cosmological
parameters that are currently affected by severe degeneracies and even
measure accurately the mass of the neutrino. The field is already in a
phase of hectic activity, data of increasing quality are being
released and many new missions devised. It is likely that if B modes
exist and their level is not surprisingly low, they will be detected
from a sub orbital effort within the next few years. However, in order
to fully exploit the information in CMB polarization we will have to
wait for a new generation satellite, that could hardly be in operation
before the 2020s. The long road to this mission will be paved by
the experiments under development.

\end{document}